\documentclass{ifacconf}

\usepackage{graphicx}      
\usepackage{cite}
\usepackage{amsmath,amssymb,amsfonts}
\usepackage{algorithm}
\usepackage[noend]{algpseudocode}
\usepackage{graphicx}
\usepackage{textcomp}
\usepackage{xcolor}
\usepackage{comment}
\usepackage{float}
\usepackage{natbib}        
\newcommand\scalemath[2]{\scalebox{#1}{\mbox{\ensuremath{\displaystyle #2}}}}
\begin{document}
\begin{frontmatter}

\title{Scenario-based model predictive control\\of water reservoir systems} 

\author[First]{Raffaele G. Cestari}, 
\author[First]{Andrea Castelletti},
\author[First]{Simone Formentin}

\address[First]{Department of Electronics, Information, and Bioengineering, Politecnico di Milano, P.za L. Da Vinci 32, 20133 Milano, Italy.\\Email to: simone.formentin@polimi.it.}

\begin{abstract}                
The optimal operation of water reservoir systems is a challenging task involving multiple conflicting objectives. The main source of complexity is the presence of the water inflow, which acts as an exogenous, highly uncertain disturbance on the system. When model predictive control (MPC) is employed, the optimal water release is usually computed based on the (predicted) trajectory of the inflow. This choice may jeopardize the closed-loop performance when the actual inflow differs from its forecast. In this work, we consider - for the first time - a stochastic MPC approach for water reservoirs, in which the control is optimized based on a set of plausible future inflows directly generated from past data. Such a scenario-based MPC strategy allows the controller to be more cautious, counteracting droughty periods (e.g., the lake level going below the dry limit) while at the same time guaranteeing that the agricultural water demand is satisfied. The method's effectiveness is validated through extensive Monte Carlo tests using actual inflow data from Lake Como, Italy.
\end{abstract}


\end{frontmatter}

\section{Introduction}
In water resources management, advanced control techniques are attracting the growing interest of researchers due to the need to adapt abruptly to droughts and extreme rainfalls. Among the available methods, \textit{Model Predictive Control} (MPC) is of great interest. It allows one to compute the optimal control action for desired objectives (e.g., flood and dry avoidance, water demand following), thus anticipating the evolution of the system and reacting promptly against sudden phenomena.

However, most of the works on the topic consider a \textit{deterministic} version of the method, namely the control sequence is optimized by considering the expected evolution of the system based on the available forecasts of the inflows, see, e.g., \cite{MPC1}, \cite{agudelo}, \cite{baunsgaard}, \cite{myanmar}, \cite{turchia}. It follows that inaccuracies in such predictions might heavily affect the final closed-loop performance. 

In this work, we propose an alternative paradigm that considers the whole distribution of the disturbance to calculate the optimal control action. The key idea is to generate a set of possible inflow scenarios starting from past realizations, leading to different system evolutions, and look at all such trajectories to compute a robust control action. We aim to show that \textit{optimizing the average performance over the distribution of the inflows might be \emph{statistically} more effective than optimizing the nominal performance for the average inflow}. We will use the scenario-based MPC approach of \cite{CampiArticle} for this purpose.

As far as we are aware, this is the first time such an approach is applied to water management systems, whereas other robust strategies (e.g. ensemble forecasts, tree-based MPC) can be found in \cite{raso}, \cite{velaverde}, \cite{ficchi}, \cite{ficchi2}. In \cite{garatti} a similar strategy is shown. The main differences are in the application, the management of a river compared to the control of a reservoir, and the use in our work of a data-driven black-box algorithm which, once trained on historical data, returns the scenario configurations. The proposed methodology is tested here on the regulation of Lake Como, a subalpine basin in northern Italy, for which past time series of the inflows are available to run a realistic simulation case study. 

The novel contributions of the paper can be listed as follows: 
\begin{itemize}
	\item application of scenario-based MPC to control the level of lake 
        Como;
	\item usage of an expanding-window strategy to forecast future inflows using a multi-seasonal multi-trend time-series model. This choice  allows us to generate a set of possible future scenarios complying with the past observational distribution;
	\item we show via extensive Monte Carlo simulations that the {scenario}-based MPC method statistically outperforms the benchmark {deterministic} strategy. {We show that a scenario strategy outperforms its deterministic counterpart by employing the same forecasting model.}
\end{itemize}
Section \ref{pb_stat} formally states the control problem. Section \ref{scenario_formulation} describes the Scenario MPC strategy, Section \ref{results} illustrates the experimental results obtained on a realistic simulator of the lake dynamics. 

\section{PROBLEM STATEMENT}

\label{pb_stat}

Without loss of generality, in this work we consider - as a case study - the predictive control of the level of lake Como, Italy. The lake is regulated to satisfy the downstream irrigation demand, while at the same time avoiding lake floods and too low-level conditions. 

Traditionally, such an MPC problem is formulated as \cite{MPC1}: 
\begin{subequations}
\label{MPCmio}
    \begin{align}
	&\scalemath{0.85}{\min_{u} 
	\,\,\sum_{t=0}^{H-1} {(s(t)-s_{max})^2+(s(t)-s_{min})^2+\lambda(u(t)-w(t))^2}}\label{cost}\\
	&s(t+1) = s(t) +3600(\hat{q}(t)-u(t))\label{sistema}\\
	&u_{min} \leq u(t) \leq u_{max}
	\end{align}
\end{subequations}
where $t$ is the time index, $H$ is the prediction horizon, $s \, [m^3]$ denotes the lake volume, $s_{max} \, (s_{min})\, [m^3]$ is the maximum (minimum) lake volume, $u \, [m^3/s]$ is the water release playing the role of the control action, $u_{max} \, (u_{min})\, [m^3/s]$ is the maximum (minimum) water release, $\hat{q}(t) \,[m^3/s]$ is the water inflow forecast, $w \, [m^3/s]$ is the agricultural water demand and $\lambda$ is a tunable hyperparameter defining the trade-off between the water demand and the level constraints.A different choice of $\lambda$ leads to different management of control objectives and system trajectories. 

From \eqref{sistema}, it is clear that the output trajectory along the prediction horizon is highly dependent on the predicted water flow $\hat{q}(t)$. Thus, as a consequence, {the quality of the controller depends on the quality of the forecast}.
Specifically, water scarcity may lead to dry periods when it is hard to satisfy the agricultural demand without decreasing excessively the lake level. On the other hand, the abundance of water might cause sudden floods that determine the relevant damages on the lake shores, particularly in the city of Como. 
For the above reasons, in the scientific literature (see, e.g. \cite{ficchi},\cite{sessa}), several hydrometeorological models have been proposed for streamflow forecasts. Nonetheless, any model would be only an approximation of the reality, and the actual disturbance might be significantly different from the predicted one.
In particular, even if forecasts are becoming more and more accurate, their prediction quality heavily depends on the considered lead time and on the seasonal period. Therefore, in this article, we do not try to improve the estimation approach, but we propose an alternative approach to deal with this source of uncertainty. Specifically, we consider a scenario-based MPC approach, in which the control action is generated by optimizing over a \textit{set} of inflow realizations, computed from an estimation of the uncertainty set made on past measurements. 
Rigorously, the control problem in \eqref{MPCmio} is reformulated as
\begin{subequations}
	\label{scenario1}
	\begin{align}
	&\scalemath{0.8}{\min_{u} \mathbb{E}\left[
	\,\,\sum_{t=0}^{H-1} {(s(t)-s_{max})^2+(s(t)-s_{min})^2+\lambda(u(t)-w(t))^2}\right]}
	\label{scenario_cost}\\
	&s(t+1) = s(t) +3600(\hat{q}(t)-u(t))	\label{scenario_sistema}\\
	&u_{min} \leq u(t) \leq u_{max}
	\end{align}
\end{subequations}
where the expected value $\mathbb{E}[\cdot]$ is computed with respect to all \textit{candidate predictors} of the inflows.

In this work, we comparatively analyse the performance of this approach with two benchmark strategies:
\begin{enumerate}
	\item a {deterministic} MPC approach, where the water inflow proxy $\hat{q}$ is defined as the \textit{daily cyclostationary average}, called \textit{climatology} (e.g., the average water inflow per day across years: 1946-1999);
	\item a deterministic MPC strategy with \textit{perfect knowledge} of the water inflow over the prediction horizon, named \textit{oracle} MPC hereafter.
        \item {a deterministic MPC approach where the nominal prediction generated by the same forecasting model used in Scenario MPC (\textit{Prophet model forecast} trained on 1999 year data) is assumed as the inflow value. This comparison is introduced in the Monte Carlo validation phase to show that the obtained performance of our robust approach is due to the scenario formulation and not to the model choice.}
\end{enumerate}

\section{Scenario-based MPC}
\label{scenario_formulation}
\begin{figure}[H]
	\centering
	\includegraphics[width=0.9 \columnwidth]{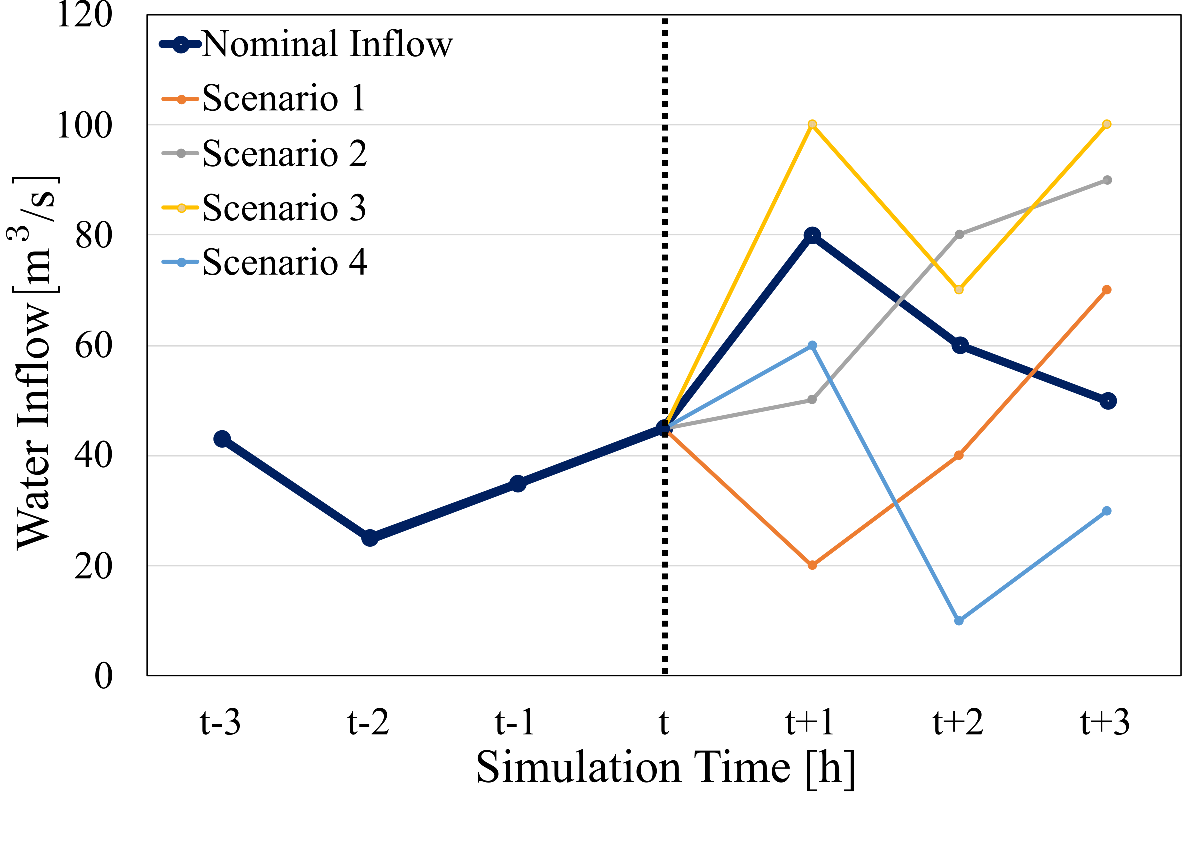}
	\caption{Graphical Representation of Lake Como Water Inflow Scenarios}
	\label{fig:distribution}
\end{figure}

The key idea of the paper is illustrated in Figure \ref{fig:distribution}: at current time $t$, the past information generates a set of candidate future realizations (e.g. \textit{scenarios}), not just the nominal one, depending on the uncertainty of the estimate. A robust control should then be optimized over such a range of disturbance values.
The robust control problem in \eqref{scenario1} can then be rephrased as:
\begin{subequations}
	\label{scenario2}
	\begin{align}
	&\scalemath{0.73}{\min_{u} \frac{1}{K} \cdot \sum_{k=1}^{K}\left[
	\,\,\sum_{t=0}^{H-1} {(s_k(t)-s_{max})^2+(s_k(t)-s_{min})^2+\lambda(u(t)-w(t))^2}\right]}
	\label{scenario_cost}\\
	&s_k(t+1) = s_k(t) +3600(\hat{q}_k(t)-u(t)) \label{scenario_sistema}\\ 
	&u_{min} \leq u(t) \leq u_{max}
	\end{align}
\end{subequations}
where the expected value operator is substituted with the statistical average and $K$ denotes the considered number of scenarios.

\subsection{Choice of $K$}
\label{sceltaK}
$K$ must be chosen large enough to achieve statistical guarantees, as indicated in \cite{CampiArticle}.
In particular, $K$ can be computed as
\begin{equation}
	K = \frac{2}{\epsilon}(ln\frac{1}{\beta}+H),
\end{equation}
where $\epsilon$ is the percentage of chance constraints that is tolerable violating, $\beta$ is the probability of having an infeasible solution to the optimization problem, and $H = 24$ is the prediction horizon length. 
\begin{figure}[htbp]
	\centering
	\includegraphics[width=1 \columnwidth]{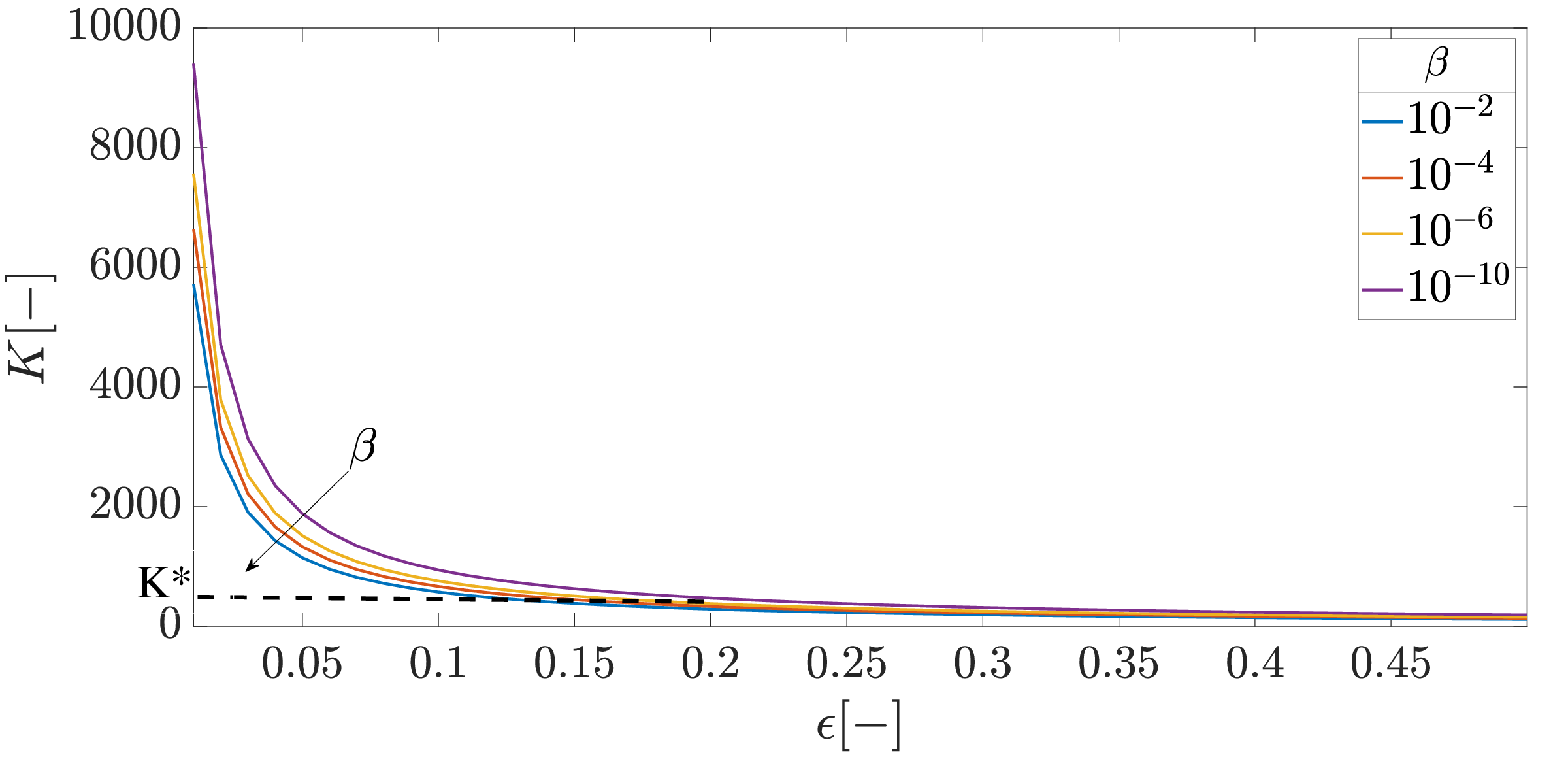}
	\caption{Number of scenarios $K$ depending on $\beta$ and $\epsilon$ }
	\label{fig:nbetaepsilon}
\end{figure}
As shown in Figure \ref{fig:nbetaepsilon}, the number of required scenarios increases as $\beta$ and $\epsilon$ decrease. More restrictive theoretical guarantees on disturbance uncertainty mandate a higher $K$.  To achieve a satisfactory theoretical bound on uncertainty, the chosen parameters are $\beta = 10^{-6}$ and $\epsilon = 0.2$, leading to $K^* = 380$. 

Table \ref{tb:Ktime} shows the practical feasibility of the \textit{scenario} strategy in the case of strong theoretical requirements, pushing towards $0$ the chances of infeasibility. The required $K = 9500$ is still affordable in a real-use case. The corresponding required computational time would be $900 \, s = 15 \, min$ (while the sampling time is of $1 \,h$). However, in this study, we will show that satisfactory performance is achieved also with a less severe setting, summarized in Table \ref{tb:Ktime}.
\begin{table}[htbp]
	\begin{center}
		\caption{Number of Scenarios vs Computational Time}\label{tb:Ktime}
		\begin{tabular}{|p{1cm}|p{1.5cm}|p{1.5cm}|p{0.8cm}|}
			\hline
			&Ideal Case &Case Study \\ \hline
			$\textbf{K} \,\,[-]$ & 9500 & 380 \\
			\hline
			\textbf{Time} $[s]$ & $~900$ & $~40$ \\
			\hline
			$\beta \,\,[-]$ & $10^{-10}$ & $10^{-6}$ \\
			\hline
			$\epsilon \,\,[-]$ & $10^{-3}$ & $0.2$ \\
			\hline
		\end{tabular}
	\end{center}
\end{table}
\subsection{Scenario generator}
\label{scenarioGenerator}
Given $K^*$, we need to select a strategy to generate the scenario set. In this work, we employ the Prophet model, an additive 
model with interpretable parameters identified by fitting non-linear trends and multiple seasonalities developed by Meta. Originally, it was developed to forecast event generation on Meta social platflorms, however, its flexible formulation helps in spreading its deployment in several applications. Its implementation is based on the underlying Stan code, a probabilistic
programming language for statistical inference written in C++, see \cite{stan}. 
Prophet looks very suitable as a \textit{scenario generator} for two main reasons: 
\begin{enumerate}
	\item it generates a \textit{distribution of possible realizations} based on the historical time-series. 
	\item it has been developed to manage at best time-series with evidence of \textit{seasonal} and \textit{trend} components, both characterizing the water inflow. 
\end{enumerate}
\begin{rem} We stress here that we are \textit{not} stating that the employed forecasting model is the best one. Our objective is instead to highlight how a predictive model embedded with a confidence interval can be used to generate a distribution of possible outcomes, and the latter can be employed within a robust control framework to outperform the existing deterministic benchmarks.
\end{rem}

Formally, the Prophet model for prediction of the nominal time-series is: 
\begin{equation}
	y(t) = g(t) +s(t) + h(t) +e_t,
	\label{prophet}
\end{equation}
where $y(t)$ is the nominal prediction, $g(t)$ is the identified trend, $s(t)$ is the seasonal component, $h(t)$ is the holiday effect component, $e_t$ is the error term. The trend $g(t)$ is defined as the piece-wise linear model
\begin{equation}
	g(t) = k(t)t+m(t),
	\label{trend}
\end{equation}
where $k(t)$ is the growth rate and its evolution through time is described by $k(t) = k_0+a(t)^T\delta$, $k_0$ being the initial growth rate and 
\begin{equation}
	a(t) = [a_1(t),...,a_j(t),...,a_M(t)]^T.
\end{equation}
$M$ is the number of breakpoints of the piece-wise linear trend, while each vector component is 
\begin{equation}
	a_j(t) = 
	\begin{cases}
		1 \, \, \, t = t_j\\
		0 \, \, \, t \neq t_j\\
	\end{cases}\,\,\, \forall j = 1,...,M
\end{equation}
Each component is $1$ only if the current time instant coincides with its associated breakpoint. Finally, 
\begin{equation}
	\delta = [\delta_1(t),...,\delta_j(t),...,\delta_M(t)]^T 
\end{equation}
is the vector of growth rate variation corresponding to each breakpoint. 
$m(t)$ is the offset parameter and its evolution in time is written as $m(t)= m_0+a(t)^T\gamma$, where $m_0$ is the initial offset parameter and $\gamma = [\gamma_1(t),...,\gamma_j(t),...,\gamma_M(t)]^T$ is a vector of offset parameter variation corresponding to each breakpoint. 
The seasonality component is defined as a Fourier series, identifying $2N$ harmonic coefficients $a_n$ and $b_n$, $P$ is the period and $N$ is the maximum harmonic order. 
\begin{equation}
	s(t) = \sum_{n=1}^{N}(a_ncos\frac{2\pi nt}{P}+b_nsin\frac{2\pi nt}{P})
	\label{season}
\end{equation}
The holiday component $h(t)$ is set to $0$ in this study. For the interested reader the complete formulation is available in \cite{prophetArticle}. 
The uncertainty distribution of the forecasted time-series is estimated using Monte Carlo simulation. By default, Prophet returns uncertainty built upon \textit{trend} and \textit{observation noise}. It assumes that \textit{the future will see the same frequency and magnitude of rate changes as the past}. 
A scenario matrix $[S]$ of size $H\times K^*$ is then returned and passed to the scenario-based MPC strategy to perform the optimization step.

\subsection{Scenario-based control}
\label{scenarioController}
In vector form, the cost function of \eqref{scenario2} becomes
\begin{equation}
	V(u,q^k) = c(||s_{max} - \textbf{s}^k||_2+||\textbf{s}^k-s_{min}||_2)+||\textbf{u}-\textbf{w}||_2
\end{equation}
where $\textbf{s}^k \, [m^3]$ is the vector of size $H$ of the lake volume evolution according to the $k^{th}$ scenario. $\textbf{w} \, [\frac{m^3}{s}]$ is the vector of size $H$ of the agricultural water demand, 
$\textbf{u} \, [\frac{m^3}{s}]$ is the vector of size $H$ of the control action, $c$ is a scaling factor to guarantee the same order of magnitude of the cost terms.
This convex cost summarizes the main Lake Como control objectives, as they are defined in \cite{MPC1}:
\begin{enumerate}
	\item \textit{flood control}: to prevent lake Como level from rising above a relative flood threshold $s_{max} \, [m^3]$.
	\item \textit{low-level avoidance}: to prevent lake Como level from going below a relative dry threshold $s_{min} \, [m^3]$.
	\item \textit{water demand}: to satisfy the downstream agricultural water demand $w\,  \frac{[m^3]}{s}$ throughout the year.
\end{enumerate} 
At each simulation step, the optimization problem becomes
\begin{subequations}
	\begin{align}
	\label{MPC}
	&\scalemath{0.855}{{\min}_{u} 
		\,\, \frac{1}{K^*}\sum_{k=1}^{K^*} {c(||s_{max} - \textbf{s}^k||_2+||\textbf{s}^k-s_{min}||_2)+||\textbf{u}-\textbf{w}||_2}}\\
	&s^k(t+1) = s(t) +3600(q^k(t)-u(t)) \\
	&u_{min} \leq u(t) \leq u_{max}\\
	&\textbf{s}^k = [s^k_1,...,s^k_{H}] \\
	&\textbf{u} = [u_1,...,u_{H}] \\
	&c = 10^{-4}, \,\, H = 24,\,\, K^* = 380
	\end{align}
\end{subequations}
where $q^k(t)$ is the $k^{th}$ water inflow scenario at time $t$.
Since each cost term is defined as a 2-norm, the optimization problem is convex and thus computationally light to solve. However, we remark this is not quadratic because the terms are not squared. This avoids excessively large values that could lead to numerical issues. 

The resulting scenario-based MPC solution is described in Algorithm \ref{alg:MYALG}.

\begin{algorithm}[H]
	\caption{Scenario MPC Algorithm}
	\label{alg:MYALG}
	\begin{algorithmic}[1]	
		\State Initialize Prophet training window: 
		$W= W_0$. 
		\For{t = 1,...,T}
		\State Enter {\textbf{Scenario Generator Module}}
		\State Read data $D$ over $W$ from Historical DB.
		\State Train Meta Prophet over $D$. 
		\State Return scenario matrix $[S]_{H \cdot K^*}$
		\State Enter {\textbf{Controller  Module}}
		\State Optimization step of Equation \eqref{MPC} over $[S]_{H \cdot K^*}$
		\State Return optimal control action $u^*(t)$.
		\State Execute Lake Como dynamics.
		\State Expand Prophet training window: $W = [W,t]$
        \EndFor
        \textbf{End for loop.}
	\end{algorithmic}
\end{algorithm}
For each simulation time step $t$ until $T$, the scenario generator is trained over the historical data $D$ and returns the scenario matrix $[S]_{H \cdot K^*}$. The control block optimizes over $[S]$ and generates the optimal control action $u^*(t)$. At each time step, the training data window $W$ (initialized using one year $W_0$: 01-01-1999 - 31-12-1999) is \textit{expanded}, to include the new observed historical data. This allows the Prophet model to always be updated to the latest data available, making the most of the data available in training to produce an accurate model.

\section{Simulation results}
\label{results}
In this section, numerical simulations are shown. Section \ref{nominale} illustrates the performance of the 3 MPC strategies with the nominal inflow. In Section \ref{scenariProphet}, 50 artificially generated scenarios are used to statistically validate the performance. {In this section, the deterministic MPC strategy using the Prophet nominal forecast is introduced}. In both sections, the simulation horizon is $T = 2160 \,h$, 3 months, starting from 01-01-2000. The Prophet model is trained (both in the scenario MPC and the deterministic MPC) starting from 1999. 
All computations are carried out on an Intel Core i7-8750H with 6 cores, at 2.20 GHz (maximum single core frequency: 4.10 GHz), with 16 GB RAM and running Matlab R2019b. The control system runs in Matlab-Simulink, the MPC optimization steps are implemented using CVX with the SDPT3 solver, and scenario generation via Prophet is implemented in Python 3.7.13.

\subsection{Nominal water inflow}
\begin{figure*}[tbph!]
	\centering
	\includegraphics[width=1 \textwidth]{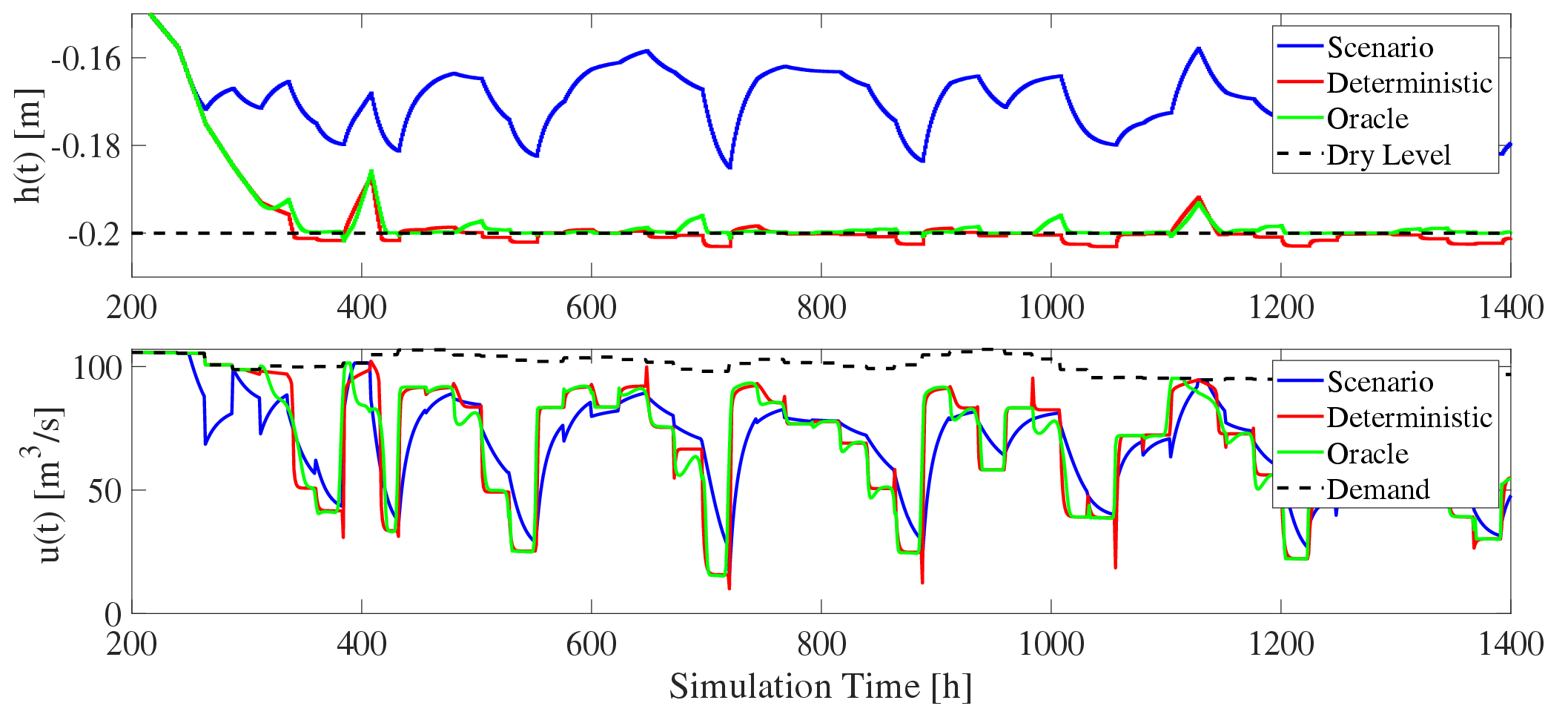}
	\caption{Lake level and water release in case the real inflow coincides with the nominal one.}
	\label{fig:true_inflow_level_release_zoom}
\end{figure*}

In the upper plot of Figure \ref{fig:true_inflow_level_release_zoom}, the lake level $h$ obtained with the 3 MPC strategies is illustrated. 
The level obtained with the {scenario}-based MPC is higher than the others. This is caused by a more \textit{cautious} behavior characterizing the scenario MPC. Many of the $K^* = 380$ scenarios are close to $0$ because the Prophet model was identified within a {dry} period with water scarcity. Therefore, the computed optimal control action is {conservative}, as seen in the lower plot, particularly around the $300^{th}$ hour. Considering the \textit{oracle} MPC (green) and the \textit{deterministic} MPC (red), the controlled variables are similar, with some relevant differences. While \textit{oracle} can maximize water release simultaneously keeping the lake level at the dry level, \textit{deterministic} tries to mimic the same behavior but poorly. The latter computes an optimal control action for its inflow knowledge, the climatology, which is higher than the true one in the simulated year.
\label{nominale}
A synthetic evaluation of the 3 methods can be obtained by defining a unique performance metric, e.g., like the following:
\begin{subequations}
	\label{non-linear cost}
	\begin{align}
		&J(t) = c_d \cdot C_{dry}(t) +C_{dmd}(t) \label{non-linear costa}\\
		&C_{dry}(t) =  
		\begin{cases}
			h_D - h(t) \, \, \, h_D > h(t)\\
			0 \, \, \, otherwise\\
		\end{cases} \\
		&C_{dmd}(t) =  
		\begin{cases}
			w(t) - u(t) \, \, \, w(t) > u(t)\\
			0 \, \, \, otherwise\\
		\end{cases}
	\end{align}
\end{subequations}
where $c_d= 10^3$ is a scaling factor needed to merge the two costs ($C_{dry}(t) ~ cm \propto 10^{-2}$, $C_{dmd}(t) ~ 10  \,\, \frac{m^3}{s} \propto 10$).
Given the simulation time period, the two main cost components are the \textit{dry lake level condition avoidance} and \textit{water demand following}. The \textit{flood avoidance} term is negligible because in the given time period it is not experienced. Thus, a cost comprising only the first two is defined. Particularly, it is constructed in a non-linear fashion to be \textit{more representative} of the lake regulation goals. Indeed, they would be better captured through \textit{if-then-else} conditions (e.g., the dry avoidance cost should appear only if the lake level is below the dry level, the water deficit cost should appear only if the water release is below agricultural demand) but they could not be used in this form within the optimization problem because they would cause non-convexity. However, they can be expressed in this fashion \textit{a-posteriori} to compare the control strategies.
\begin{figure}[htbp]
	\centering
	\includegraphics[width=1\columnwidth]{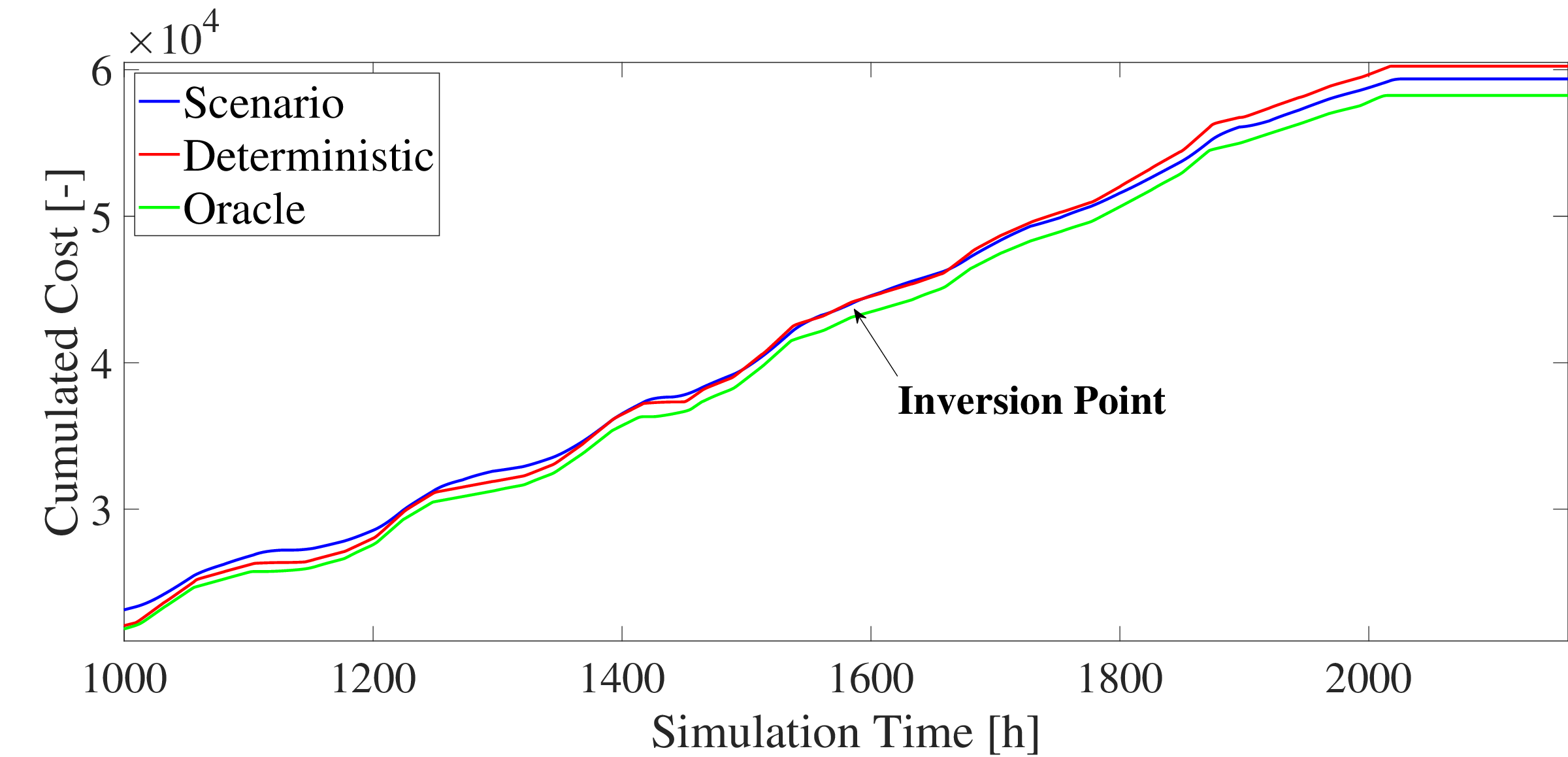}
	\caption{Cumulated nonlinear cost - the case of nominal inflow.}
	\label{fig:non_linear_cost_10_3}
\end{figure}
In Figure \ref{fig:non_linear_cost_10_3} the cumulated version of \eqref{non-linear costa} is shown. Two main observations can be made: 
\begin{enumerate}
	\item the terminal cumulated costs show that the strategy with the highest penalty is the {deterministic} one, followed by {scenario} and {oracle}. As expected, {oracle} has the best performances. The {scenario-based} strategy, although there were no a priori theoretical guarantees, shows lower cost with respect to the {deterministic} one.
	
	\item among the cost curves, a \textit{switch} between the {deterministic} (red) and the {scenario-based} (green) is observed. This is of particular interest since it means that initially, the cumulated cost of the deterministic strategy is lower. This comes from the fact that the cyclostationary average is more representative of the nominal inflow in that period. Later on, the red curve overlaps and then overcomes the blue one, indicating a sequence of cumulated errors which lead to better performances of the scenario-based strategy. 
\end{enumerate}
\begin{table}[htbp]
	\begin{center}
		\caption{Performance of MPC strategies: dry avoidance and water demand tracking.}\label{tb:nl_cost}
		\begin{tabular}{|p{3.4cm}|p{0.95cm}|p{1.7cm}|p{0.9cm}|}
			\hline
			&Scenario &Deterministic &Oracle\\ \hline
			Minimum Lake Level $[m]$ & -0.185 & -0.205 & -0.202 \\
			\hline
			Dry Hours & 0 & 1316 & 41 \\
			\hline
			Demand Deficit Peak $[m^3/s]$ & -86.4 & -92.4 & -82.9 \\
			\hline
			Demand Deficit Hours & 1974 & 1941 & 1960 \\
			\hline
		\end{tabular}
	\end{center}
\end{table}
Table \ref{tb:nl_cost} numerically explains the conclusions achieved by looking at Figure \ref{fig:non_linear_cost_10_3}:
\begin{enumerate}
	\item The {scenario-based} solution never fails with respect to the dry objective. 
	\item The \textit{oracle} solution has very low error on dry objective (only $41$ hours of dry condition).
	\item The {deterministic} performs poorly  with respect to the dry objective, over 1300 hours of dry condition. 
	\item The lowest deficit peak on water demand following is achieved by the {oracle} algorithm. 
	\item The highest number of water demand deficit hours is achieved by the {scenario-based} strategy, depending on its cautelative policy. The lowest number of deficit hours is achieved by the {deterministic} MPC at the cost of the issue discussed at the previous bullet (2).
\end{enumerate}

\subsection{Prophet-based water inflows generation}
\begin{figure}[H]
	\centering
	\includegraphics[width=0.8\columnwidth]{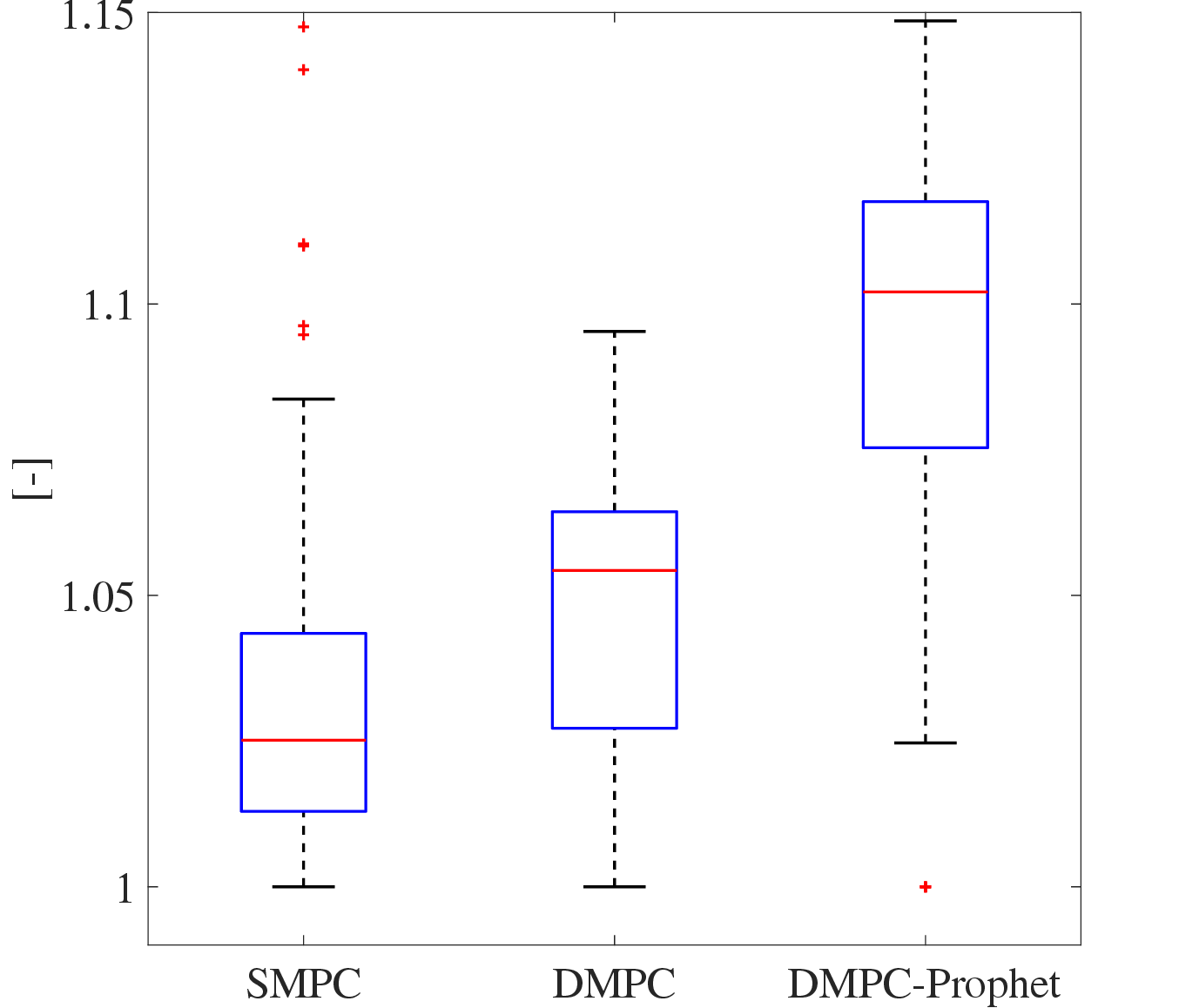} 
	\caption{Normalized costs - 50 validation scenarios.}
	\label{fig:cumulated_50_scenarios}
\end{figure}

\begin{figure}[H]
	\centering
	\includegraphics[width=1\columnwidth]{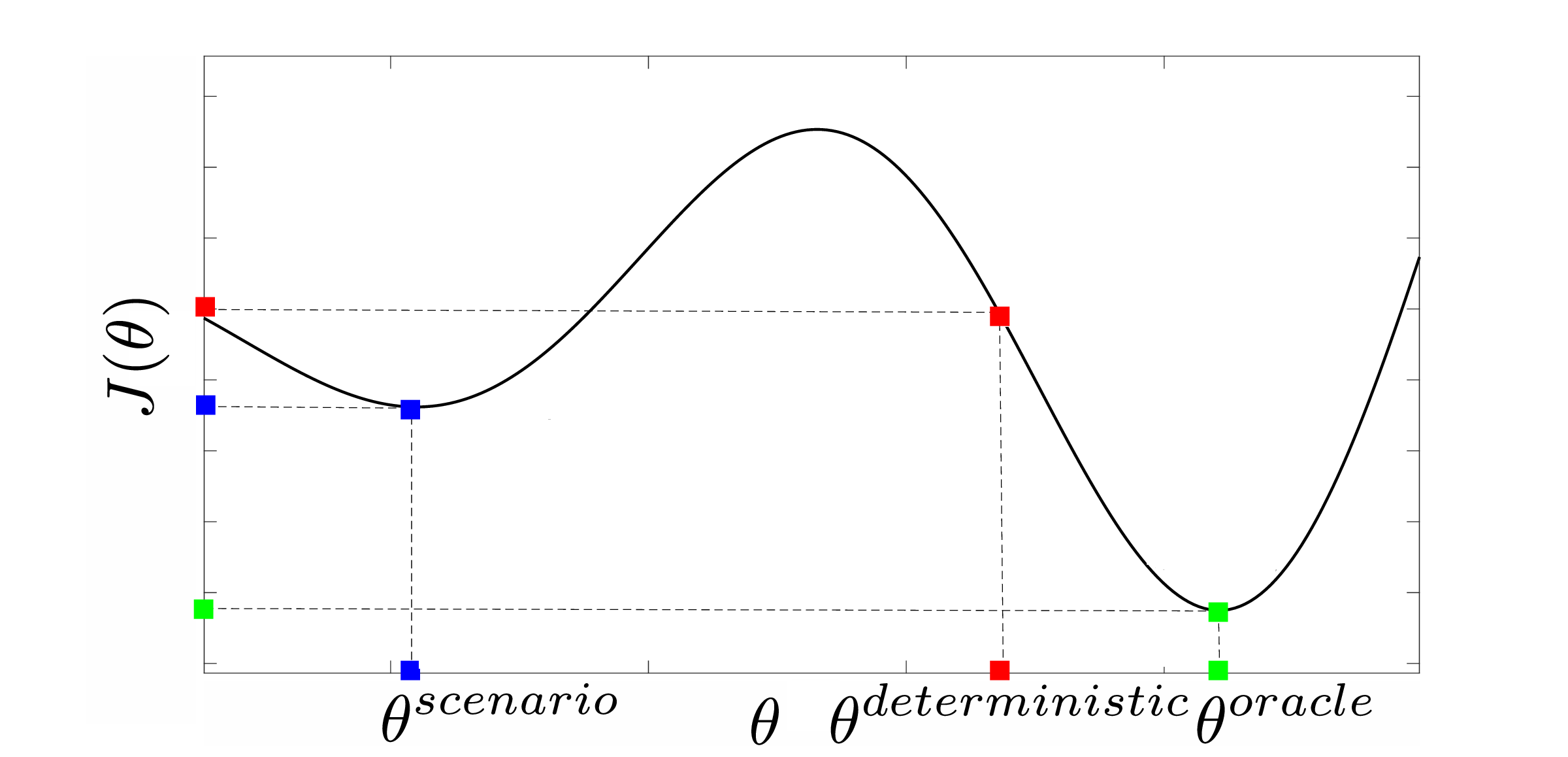}
	\caption{Control cost $J(\theta)$ as function of control strategy parameters $\theta$.}
	\label{fig:parametri}
\end{figure}

This section shows the statistical validation of the \textit{scenario} strategy on $50$ scenarios generated by Prophet, trained on the historical time-series. Although we did not have any theoretical guarantee about the effectiveness of the \textit{scenario} approach on specific realizations, it is evident how the technique can, statistically, overcome the \textit{deterministic} strategy. The proof is shown in Figure \ref{fig:cumulated_50_scenarios}. Each box collects the cumulated non-linear costs computed according to Equation \eqref{non-linear costa} of {3 strategies: \textit{scenario} (SMPC) \textit{deterministic-cyclostationary} (DMPC) and \textit{deterministic-prophet} (DMPC-Prophet)}. Each cost corresponds to a specific realization of 1 out of the 50 scenarios, normalized to the \textit{oracle} cost obtained in the same scenario. This normalization is done because \textit{oracle} experiences \textit{always} the lower cost, and for this, \textit{oracle} boxplot is not shown. On average, \textit{scenario} performance is better than \textit{deterministic}, {both considering cyclostationary and Prophet forecasts. This allows us to conclude that, although Prophet model might not be the best possible choice for this application, indeed, benchmark deterministic MPC with cyclostationary average outperforms deterministic MPC with Prophet nominal forecast, \textit{scenario} approach is outperforming. Thus, we have successfully isolated the scenario-based control strategy's impact on managing the considered multi-objective control problem, proving its statistical effectiveness}.
However, as the outliers remarked as red crosses highlight, it might happen that, for specific realizations, \textit{deterministic} approaches work better. This depends on the closeness of the given scenario to the deterministic forecasts. Figure \ref{fig:parametri} shows a qualitative representation of the behavior of the 3 control strategies. Although the shape of the control response of \textit{deterministic} is similar to \textit{oracle}, it does not necessarily correspond to a lower control cost. It is shown that the cost obtained with the \textit{scenario} is lower than that of the \textit{deterministic}. However, the parameters associated with the strategy are less close to \textit{oracle}. The proximity of the control policies exercised by \textit{oracle} and \textit{deterministic} controllers is not reflected in proximity in the corresponding control costs, depending on the \textit{failure} of finding the corresponding global minima by \textit{deterministic}.
\vspace{-7pt}
\label{scenariProphet}
\vspace{-16pt}
\vspace{-10pt}
\section{Conclusions}
In this work, we present a scenario-based MPC solution for water resources management and we apply it to the regulation of Lake Como, Italy. {A numerical comparison with 3 deterministic MPC methods, one using the full knowledge of the inflow (\textit{oracle} MPC), one employing an approximation of the disturbance realization (DMPC) based on benchmark cyclostationary average and one using the Prophet nominal forecast, (DMPC-Prophet)}, showed that our \textit{scenario} MPC technique could be an effective tool to solve the multi-objective real-time control problem within this domain, robustly against upcoming increasingly dry future climates. Notwithstanding its computational load, considering the sampling time of the application (1 hour), the strategy can be effectively used in practice with appropriate probabilistic guarantees. Future work includes the analysis of the scenario-based solution using different scenario generators, e.g., models learned via kernel-based methods that supply an estimate of the uncertainty region together with the nominal forecast.
\bibliography{ifacconf}             
                                                   






\end{document}